\documentclass[twocolumn,amsmath,amssymb]{revtex4}

\usepackage{graphicx}
\usepackage{bm}

\begin{document}

\title{Incident-Energy Dependence of the \\ 
Effective Temperature in Heavy-Ion Collisions}

\author{M. Ga\'zdzicki$^a$, 
M. I. Gorenstein$^b$, 
F. Grassi$^c$, Y. Hama$^c$, T. Kodama$^d$ and
O. Socolowski Jr.$^c$}
 
\affiliation{
$^a$ Institut f\"ur Kernphysik, Universit\"at Frankfurt, Germany. \\
$^b$ Bogolyubov Institute for Theoretical Physics, Kiev, Ukraine. \\
$^c$ Instituto de F\'{\i}sica, Universidade de S\~ao Paulo, CP 66318, 
S\~ao Paulo, 05315-970, SP, Brazil. \\
$^d$ Instituto de F\'{\i}sica, Universidade Federal do Rio de Janeiro, CP 68528,
Rio de Janeiro, 21945-970, RJ, Brazil.
}

\date{\today}%

\maketitle

\section{Introduction}

Experimental data on transversal momentum distributions 
of kaons produced in central Pb+Pb or Au+Au collisions show an anomalous 
dependence on the collision energy \cite{ags, sps, rhic}. The effective temperature $T^*$, 
or slope parameter of 
the transversal-momentum spectra~\cite{note1}, 
increases with energy in AGS and RHIC energy domains. 
However, in the SPS energies the effective temperature keeps 
approximately constant.  Recently, it was argued~\cite{MG2a} that this behaviour 
might be caused by a modification of 
the equation of state in the transition region between confined and deconfined matter,
as suggested by Van Hove~\cite{Hov} a long time ago, 
and 
could be considered as new signal of deconfinement in the SPS energy domain.

Our main object in the present work is to 
study the behaviour of the effective temperature 
for K$^+$ in several energy domains. For this purpose, we apply 
the recently developed SPheRIO~\cite{Aguiar2001,Aguiar2002} code for hydrodynamics 
in 3+1 dimensions, 
using both Landau-type compact initial conditions and 
spatially more spread ones. For the latter we used the average over many events 
generated by NeXus~\cite{nexus1, nexus2}. 
We show that initial conditions given in 
small volume, like Landau-type ones, are unable to reproduce the effective temperature 
together   with other data (multiplicities and rapidity distributions). These quantities 
can be reproduced altogether only when using a large initial volume with an appropriate velocity 
distribution.

Besides, it seems that the increase of $T^*$ in the RHIC energy domains is caused 
mainly by the larger expansion time in the hadronic phase for higher incidente energy, 
which implies lower freeze-out temperature as $\sqrt{s}$ increases~\cite{HN, NNOP}. 
It seems that, within our analyses with NeXus initial conditions, the RHIC incident energies
are not enough to produce noticeable increase in the transverse acceleration during the 
quark-gluon plasma (QGP) phase.

\section{Hydrodynamics in 3+1 dimensions}

The hydrodynamical  code in 3+1 dimensions used here
is based on the technique called 
\emph{Smoothed Particle Hydrodynamics} (SPH)~\cite{Lucy1977, Gingold1977}. This is 
a method which uses Lagrangian coordinates. 
The fluid is represented by small volumes called SPH-particles and the equations for
the fluid evolution  become a system of ordinary differencial equations for the 
SPH-particles,  
in this representation. 

The numerical code SPheRIO~\cite{note2} 
is a suitable implementation of this
method  for 
the relativistic nuclear collisions. An advantage of this technique 
is the convenience in the study of problems where the geometry  is highly irregular, 
as is the case of non-central relativistic collisions. A SPH-particle has attached to it  
conserved quantities; in the present version of SPheRIO, the entropy and the baryonic number 
are the quantities 
which are kept constant during the fluid evolution. In the SPH representation, the entropy density 
and the baryonic density are parametrized as: 
\begin{eqnarray}
s(\tau,\bm{x}) &=& \frac{1}{\tau\gamma}\sum_i\nu_i^s W(\bm{x}-\bm{x}_i; h), \\
n_b(\tau,\bm{x}) &=& \frac{1}{\tau\gamma}\sum_i\nu_i^b W(\bm{x}-\bm{x}_i; h), 
\end{eqnarray}
where $\nu_i^s$ ($\nu_i^b$) is the entropy (baryonic number) of the $i$-th SPH-particle, 
$\gamma=1/\surd(1-v_x^2-v_y^2-\tau^2 v_\eta^2)$, and $W$ is the 
interpolating kernel with 
width $h$. Here we use the hyperbolic coordinates $\tau\equiv\surd{(t^2-z^2)}$, 
$x$, $y$ and $\eta\equiv1/2 \cdot \ln[(t+z)/(t-z)]$ which are convenient for a system in rapid 
longitudinal expansion. The equations for the SPH-particles are given by:
\begin{widetext}
\begin{eqnarray}
\frac{d\bm{x}_i}{d\tau}&=&\bm{v}_i, \\
\frac{d}{d\tau}\left(\gamma\nu^s\;\frac{P+\varepsilon }{s}\;\bm{\omega }\right)_i &=&
-\sum_j \nu_i^s \nu_j^s\left[\frac{P_i}{\tau\gamma_i^2 s_i^2}+
\frac{P_j}{\tau\gamma_j^2 s_j^2}\right] \bm{\nabla}_i W(\bm{x}_i-\bm{x}_j; h),
\end{eqnarray}
\end{widetext}
where $\bm{\omega}=(v_x,v_y,\tau^2 v_\eta)$. The pressure, $P$, and the energy density, 
$\varepsilon$, are related to the entropy density $s$ and the baryonic density $n_b$  
via equation of state.

\subsection{Decoupling criterion}

The distribution of final particles is obtained by using Cooper-Frye's 
prescription~\cite{CF}:
\begin{equation}\label{eq:cf}
E\frac{d^3N}{dp^3}=\int_{\Sigma _{fo}} d\sigma \cdot p f(p\cdot u),
\end{equation}
where $f(p\cdot u)$ is the distribution function (Bose-Einstein or Fermi-Dirac) 
and the integral is evaluated on the freeze-out surface $\Sigma _{fo}$. 
In the SPH representation,  Eq. (\ref{eq:cf}) is written as
\begin{equation}
E\frac{d^3N}{dp^3}=\sum_i 
\frac{\nu_i^s\; \hat{n}_i\cdot p}{s_i\; |\hat{n}_i\cdot u_i|} 
f(p\cdot u_i),
\end{equation}
where the sum is over all SPH-particles on the freeze-out surface and 
$\hat{n}$ is the normal four-vector to this surface, 
given by $\hat{n}_\mu\propto (-\partial T/\partial \tau, -\partial T/\partial x, 
-\partial T/\partial y, -\partial T/\partial \eta)$.

\section{Equations of state}

Let us consider here two sets of equations of state. 
The first one type-I (EOS-I) has a first-order phase 
transition between an ideal gas of massless quarks (u, d and s) and gluons and an ideal 
hadron gas (baryon 
number is assumed to be zero), where the pressure, the energy density and 
the entropy density  are given by:
\begin{eqnarray}
P_q &=& \frac{\pi^2}{90}g_q T^4-B, \\ 
\varepsilon_q &=& \frac{\pi^2}{30}g_q T^4+B, \\
s_q &=& \frac{2\pi^2}{45}g_q T^3, \\
P_h &=& \frac{\pi^2}{90}g_h T^4, \\
\varepsilon_h &=& \frac{\pi^2}{30}g_h T^4, \\
s_h &=& \frac{2\pi^2}{45}g_h T^3.
\end{eqnarray}
Here $g_h=16$ is an effective parameter (see~\cite{MG2b} for details), $g_q=95/2$, and 
$B$ is the bag model parameter.

The second one, type-II (EOS-II), somewhat more realistic than the previous one,  
considers a first-order phase transition between a 
QGP
and a hadronic resonance gas 
(baryon number is taking into account). In the QGP we consider an ideal gas 
of massless quarks (u, d, s) and gluons. The thermodynamical 
quantities are given by
\begin{eqnarray}
P_q &=& \frac{\pi^2}{90} g_q T^4+
\frac{3}{2}\left(T^2\mu_q^2+\frac{\mu_q^4}{2\pi^2}\right)-B, \\
\varepsilon_q &=& \frac{\pi^2}{30} g_q T^4+
\frac{9}{2}\left(T^2\mu_q^2+\frac{\mu_q^4}{2\pi^2}\right)+B, \\
s_q &=& \frac{2\pi^2}{45} g_q T^3+
3 T\mu_q^2, \\
n_q &=& 3 \left(T^2\mu_q^2+\frac{\mu_q^3}{\pi^2}\right).
\end{eqnarray}
The hadronic phase is composed of 
resonances with mass below 2.5 GeV/c$^2$, 
where  volume correction is taken into account. 
The corrected-volume pressure, $P^{ex}$, is written in function of the pressure 
for an ideal gas $P^{id}$:
\begin{equation}
\sum_i P_i^{ex}(T,\mu_i)=\sum_iP_i^{id}(T,\tilde{\mu}_i),
\end{equation}
where $\tilde{\mu}_i=\mu_i-V_i\sum_jP_j^{id}(T,\tilde{\mu}_j)$, 
($V_i$: volume of the hadron $i$). Others thermodynamical quantities are given 
following the relation:
\begin{equation}
\chi_i^{ex}(T,\mu_i)=\frac{\chi_i^{id}(T,\tilde{\mu}_i)}{1+\sum_jV_jn_j^{id}(T,\tilde{\mu}_j)},
\end{equation}
where $\chi$ represents $n$, $\varepsilon$ or $s$.

In both equations of state,  the transition temperature is assumed to be 
160~MeV. 

\section{\label{sec:s1}Initial conditions and results}

In order to study the behaviour of the effective temperature,
we began with Landau-type initial conditions, where the fluid is at rest at $t=0$~fm/c, 
and the matter is localized in a Lorentz contracted sphere. The initial energy density 
is assumed to be constant and 
given by: $\varepsilon_0=E/V$, where $E=\zeta(\sqrt{s}-2m)A$ is the available 
kinetical energy in 
the collision, and 
$V=\gamma_0^{-1}V_0=\gamma_0^{-1}(4/3)\pi R^3$ is the 
volume of the contracted incident nuclei when superposed. 
Here $\zeta$ is the inelasticity parameter 
and $\gamma_0=\sqrt{s}/2m$.
These 
initial conditions (together with EOS-I)
can reproduce both the particle multiplicities and the rapidity 
distributions fairly well (Ref.~\cite{MG2b}). 
However it has
an inconvenience that the initial energy density is 
too high. As a consequence, the time of expansion becomes very large and transverse 
expansion is considerable. The result is a very large effective temperature and a 
non-exponential shape of the spectra in transverse momentum.

\begin{figure}
\center{\includegraphics*[width=8.8cm]{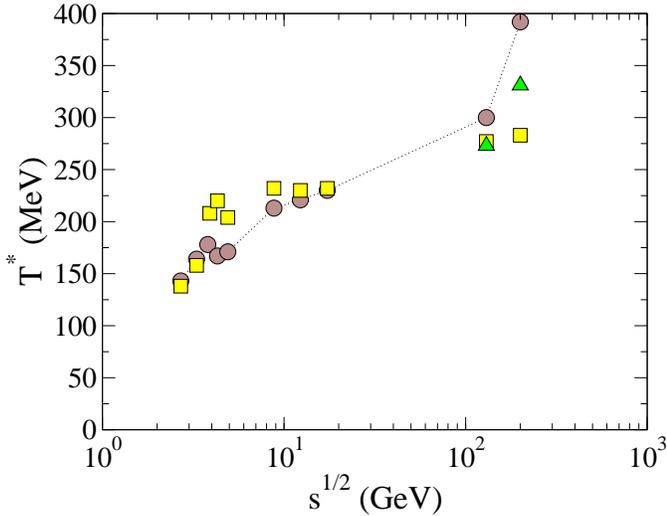}}
\caption{\label{fig:f1}Energy dependence of the effective temperature for K$^+$. 
The initial conditions are Landau-type ones, with initial longitudinal velocity. 
Triangles corresponde to $T_{fo}$=155~MeV. Experimental data are shown in square symbols.}
\end{figure}

A solution we found to this problem was 
to give the fluid an initial longitudinal-velocity distribution. We did not change the shape of 
the region where the fluid is formed, which remained a Lorentz contracted sphere. The 
initial longitudinal velocity is not constant, but proportional to $z$:
\begin{eqnarray}
u^\mu(z) &=& (\gamma(z),0,0,\gamma(z)v), \\
u^z(z) &=& \frac{z}{z_m}\,\zeta\sqrt{\gamma_0^2-1}, \\
\gamma(z) &=& \sqrt{\left(\frac{z}{z_m}\right)^2\zeta^2(\gamma_0^2-1)+1},
\end{eqnarray}
where $z_m=R/\gamma_0$.
As a consequence of this change, the initial energy density $\varepsilon_0$, 
which we took constant and 
obtained solving the equation 
\begin{equation}
E = \int\limits_{V:\;\;r^2  + \gamma _0 ^2 z^2  = R^2 } 
\left[ (\varepsilon _0  + p) \gamma ^2 (z) - p \right] dV, 
\end{equation}
became much smaller and then the transverse expansion 
compatible with data, as shown in Table \ref{tab:table1} and Fig.~\ref{fig:f1}. 
In this simulation we consider freeze-out temperatures smoothly 
increasing with energy, until SPS domains. For RHIC energies, we show 
two cases: a decreasing freeze-out temperature and a increasing one 
(bold-faced types in Table~\ref{tab:table1} and triangle in Fig.~\ref{fig:f1}).

However, since $\varepsilon_0$ became much smaller than the previous case, 
so did the entropy density $s_0$ and the total multiplicity, although the rapidity 
distributions remained more or less the same in the shape. 

\begin{figure}
\center{\includegraphics*[width=8.8cm]{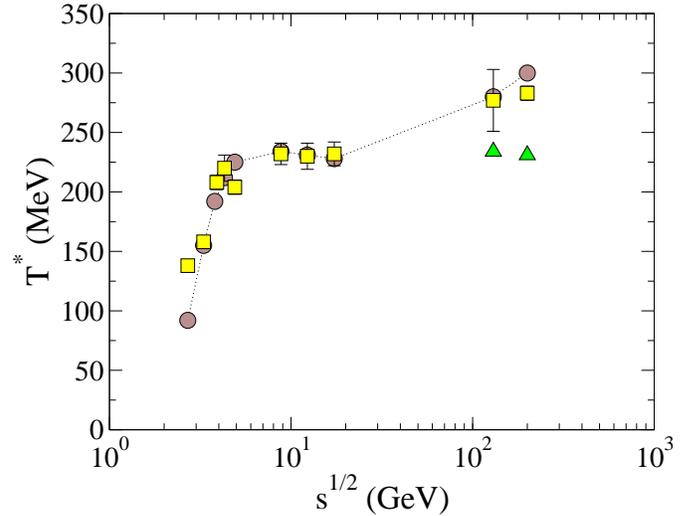}}
\caption{\label{fig:f2}Energy dependence of the effective temperature for K$^+$. 
The initial conditions are NeXus-type ones. 
Triangles corresponde to $T_{fo}$=155~MeV. Experimental data are shown in square symbols.}
\end{figure}

To reproduce all quantities (multiplicities, rapidity distributions and transverse-momentum 
distributions) we have to take a large initial volume, with some appropriate longitudinal-velocity 
distribution. To do this, we use the NeXus event generator, 
which produces the event-by-event initial conditions at time $\tau=1$~fm/c. 
Instead of using these fluctuating initial conditions, here we smooth them out by 
averaging 30 
random events for each collision energy.
For this 
case we use the EOS-II. The results are shown in Table \ref{tab:table2} and 
Fig. \ref{fig:f2}. Besides reproducing the effective temperature data quite well, 
it was shown that this initial conditions give good results also of 
multiplicity and rapidity distributions~\cite{HGSAKPTO}. 

\begin{table}
\caption{\label{tab:table1}Results obtained from Landau-type initial conditions with 
initial longitudinal velocity.  $\bar{v}_T$ is the transversal 
velocity  averaged in the rapidity interval $-0.5\le y\le 0.5$.
$\varepsilon_0$ is the initial energy density and $T_0$ is the corresponding initial temperature.}
\begin{tabular}{|c|c|c|c|c|c|}
\hline\hline
$\sqrt{s}$ &$T_{0}$ & $\varepsilon_0$&   $T_{fo}$ & $\bar{v}_T$  &  $T^*$\\ 
(A$\cdot$GeV)&(MeV)&(GeV/fm$^3$) &(MeV) & & (MeV)\\
\hline\hline
2.7	&   109	&	0.097	&	85	&          0.240     &	143	\\
\hline
3.3	&   128	&	0.185	&	94	&	0.255	&	164	\\
\hline
3.8	&   140	&	0.263	&	97	&	0.280	&	178	\\
\hline
4.3	&  149	&	0.341	&	115	&	0.155	&	167	\\
\hline
4.9	&  158	&	0.431	&	120	&	0.153	&	171	\\ 
\hline
8.8	&  160	&	1.038	&	143	&	0.146    &	213	\\ 
\hline
12.3	&  160	&	1.322	&	147	&	0.147	&	221	\\ 
\hline
17.3	&  160	&	1.536	&	149	&	0.139	&	230	\\ 
\hline
130	&  167	&	1.896	&	128	&	0.292	&	300 	\\ 
\cline{4-6}
& & &\textbf{155 }& \textbf{0.191} & \textbf{273} \\
\hline
200	&  168	&	1.908	&	125	&	0.485 	&	392	\\ 
\cline{4-6}
& & & \textbf{155} & \textbf{0.272} & \textbf{331} \\
\hline
\end{tabular}
\end{table}

\begin{table}
\caption{\label{tab:table2}Results obtained from NeXus-type initial conditions. 
The initial energy density $\varepsilon_0$ and the initial 
temperature $T_0$ are given in the central point ($x=y=\eta=0$). 
$\bar{v}_T$ is the transversal 
velocity   averaged in the rapidity interval $-0.5\le y\le 0.5$. }
\begin{tabular}{|c|c|c|c|c|c|}
\hline\hline
$\sqrt{s}$ &$T_{0}$ & $\varepsilon_0$&   $T_{fo}$ & $\bar{v}_T$  &  $T^*$\\ 
(A$\cdot$GeV)&(MeV)&(GeV/fm$^3$)& (MeV) & & (MeV) \\
\hline\hline
2.7	&   98	&	0.75	&	85	&          0.067     &	92	\\
\hline
3.3	&   128	&	0.66	&	94	&	0.28	&	155	\\
\hline
3.8	&   131	&	1.01	&	97	&	0.41	&	192	\\
\hline
4.3	&  135	&	1.38	&	115	&	0.37	&	212	\\
\hline
4.9	&  140	&	1.55	&	120	&	0.39	&	225	\\ 
\hline
8.8	&  198	&	4.06	&	143	&	0.32    &	234	\\ 
\hline
12.3	&  248	&	9.04	&	147	&	0.32	&	231	\\ 
\hline
17.3	&  265	&	11.37	&	149	&	0.32	&	228	\\ 
\hline
130	&  279	&	12.86	&	128	&	0.52	&	280 	\\ 
\cline{4-6}
& & &\textbf{155} & \textbf{0.34} & \textbf{234} \\
\hline
200	&  277	&	12.48	&	125	&	0.56 	&	300	\\ 
\cline{4-6}
& & & \textbf{155} & \textbf{0.34} & \textbf{231} \\
\hline
\end{tabular}
\end{table}

\section{Conclusions}

We conclude that
the initial conditions given in small volume like Landau-type 
discussed above are 
not appropriated to describe the multiplicity data, rapidity distributions and the 
transverse-momentum distributions at the same time. If we try to obtain the correct 
multiplicities (together with the rapidity distributions), 
the effective temperature becomes too large 
(the first case discussed in Sec.~\ref{sec:s1},  where the fluid is at rest) . 
On the other side, if we try to get the correct    
effective temperature  and the shape of rapidity 
distributions as we did, the multiplicities become small, unless an additional entropy 
is generated during the expansion (the second case discussed in Sec.~\ref{sec:s1}). 
It is clear that these conclusions about the initial conditions do not change if we replace 
EOS-I by EOS-II. 

Nexus type initial conditions, with spatially spread energy and velocity distributions, 
reproduce well all the main characteristics of data, namely particle multiplicities, 
rapidity distributions and $p_T$ spectra. In this case, the increase of $T^*$ in RHIC 
energy domain seems to be due mostly to the larger expansion in the hadronic phase 
than to that in the QGP phase.

\section*{Acknowledgments}

This work was partially supported by FAPESP  (Contract Nos. 2001/09861-1 and 2000/04422-7).


\begin{thebibliography}{99}
\bibitem{note1} The effective temperature, $T^*$, 
is obtained by fitting the transverse-mass distribution with the function 
$dN/dm_T=A\;m_T\exp(-m_T/T^*)$.
\end{thebibliography}
\end{document}